\input{aipcheck}

\documentclass[
    ,final            % use final for the camera ready runs
  ]
  {aipproc}

\layoutstyle{6x9}

\newcommand{\beq}{\begin{equation}}
\newcommand{\eeq}{\end{equation}}

\newcommand{\ord}[1]{\mathcal{O}{(#1)}}

\def\lsim{\mathrel{\raise.3ex\hbox{$<$\kern-.75em\lower1ex\hbox{$\sim$}}}}
\def\gsim{\mathrel{\raise.3ex\hbox{$>$\kern-.75em\lower1ex\hbox{$\sim$}}}}
\def\Lphi{\Lambda_\phi}

\begin{document}

\title{Signals of Warped Top-Condensation in $B$-Decays}

\classification{12.10.-g}
\keywords      {Warped top-condensation model, light radion, $B$-decays}

\author{Hooman Davoudiasl }{
  address={Department of Physics, Brookhaven National Laboratory, 
Upton, NY 11973, USA}
}

%\author{<author2>}{
 % address={<common address for author2 and author3>}
%}

%\author{<author3>}{
%  address={<common address for author2 and author3>}
 % ,altaddress={<author1 address>} % additional visiting address
%}

\begin{abstract}
 
A recently proposed warped top-condensation model generally 
predicts a light radion $\phi$, with a mass of a few GeV or less, whose 
interactions are suppressed by a scale of order 100~TeV.  In this talk, we present existing  
constraints and potential signals from the process $b \to s \phi$, 
in the context of such a model.  Astrophysical 
bounds, applicable to very light radions, are also briefly discussed.

\end{abstract}

\maketitle

%%%%%%%%%%%%%%%%%%%%%%%%%%%%%%%%%%%%%%%%%%%%
%% MAINMATTER
%%%%%%%%%%%%%%%%%%%%%%%%%%%%%%%%%%%%%%%%%%%%

The following is based on a talk given at the 
SUSY 2009 conference, held at Northeastern University, Boston, MA, USA, June 5-10, 2009.  
The material for the  talk is from Ref.~\cite{Davoudiasl:2009xz}, where more details 
and references can be found.

The warped top-condensation model (WTCM) introduced in Ref.~\cite{Bai:2008gm} is based 
on the Randall-Sundrum (RS) model \cite{Randall:1999ee}.  The WTCM does not have 
a fundamental scalar Higgs field, but includes 
bulk Standard Model (SM) gauge and fermion fields, with the latter arranged such that 
the light fields are localized towards the UV-brane and the heavy ones are localized towards 
the IR-brane of the RS background.  In particular, the top quarks are IR-localized.  

The strong interactions of the KK gluons with the top 
quarks gives rise to quark condensation, and hence 
EWSB.  If the potential for the radion scalar, associated 
with the size $L$ of the extra dimension, is dominated by top condensation, one can show that 
$k L \approx 30$, where $k$ is the 5D curvature scale.  
With $k$ near the 4D Planck mass $M_P\sim 10^{18}$~GeV, this setup 
gives rise to the appearance of the gauge Kaluza-Klein (KK) modes at $m_{KK}\sim 30$~TeV and a 
very light radion $\phi$ of mass $m_\phi \sim 1$~GeV.  The coupling of $\phi$ to the 
trace of energy-momentum tensor is suppressed by the scale
\beq
\Lambda_\phi=\sqrt{6 M_5^3/k} \, e^{-k L}, 
\label{Lamphi}
\eeq
where $M_5$ is the 5D fundamental scale of the RS background and $M_P \simeq M_5^3/k$.  
Roughly, we have 
\beq
m_\phi \sim (35~{\rm TeV}/\Lambda_\phi)\, 4{\rm~GeV}.
\label{mphi}
\eeq

The WTCM model also predicts a heavy composite Higgs of mass $m_H\sim 500$~GeV 
and requires a color-charged state 
with a mass around 2~TeV to obtain the correct top mass \cite{Bai:2008gm}.  However, here 
we are interested in the low energy signals associated with $\phi$, and in particular 
those from $B$-meson decays.  We note that 
a similar setup, using a fourth generation quark to form the EWSB condensate, was introduced in 
Ref.~\cite{Burdman:2007sx}, in which $m_{KK}\gsim 1$~TeV and $m_H\gsim 700$~GeV. 

Let us focus on the process $B \to X_s\, \phi$, with $\phi$ emitted on-shell.  We will 
use the spectator approximation: 
${\rm Br}(B \to X_{s}\, \phi)/{\rm Br}(B \to X_{c} \, e {\bar \nu_e})
\approx {\rm Br}(b \to s \, \phi)/{\rm Br}(b \to c \, e {\bar
\nu_e})$ throughout.  We begin with the loop-level contribution.  
By adapting a similar computation for a light Higgs in Ref.~\cite{Grinstein:1988yu}, we find
\beq
\frac{{\rm Br}(B \to X_{s}\, \phi)}{{\rm Br}(B \to X_{c} \, e {\bar \nu_e})}  = 
\frac{27 \sqrt{2} }{64 \pi^2}\frac{G_\phi\, m_b^2}{f(m_c/m_b)} 
\left|\frac{V_{st}^\dagger V_{tb}}{V_{cb}}\right|^2
\left (\frac{m_t}{m_b}\right )^4 
\times \left (1 - \frac{m_\phi^2}{m_b^2}\right)^2, 
\label{loopform}
\eeq
where $G_\phi \equiv (\sqrt{2} \Lphi^2)^{-1}$, and
$f(m_c/m_b)\simeq 0.5$.  With $m_\phi \sim 1$~GeV and $\Lphi \sim100$~TeV, we find
\beq
{\rm Br}(B \to X_{s}\, \phi) \sim 10^{-5}\; \; \;\;({\rm Loop}).
\label{loopnum}
\eeq

Here, we note that bulk fermions have non-universal couplings to the radion 
\cite{Azatov:2008vm}.  This gives rise 
to tree-level flavor changing neutral current (FCNC) effects that could mediate $b \to s \phi$.  The 
radion-mediated FCNC interctions, after mass matrix diagonalization, are given by \cite{Azatov:2008vm}
\beq
{\cal L}_{FV} = \frac{\phi}{\Lphi}({\bar d}^i_L d^j_R a_{ij} \sqrt{m_i m_j} + {\rm h.c.}) \; ;
\quad (i\neq j),
\label{FV}
\eeq 
where $a_{ij}$ are functions of 5D fermion profile parameters $c_f\equiv m_5^f/k$, with $m_5^f$ 
the 5D fermion mass, and $m_i$ are the SM fermion masses.

We find $a_{ij} \sim 0.05$ to be 
a good representative value \cite{Azatov:2008vm}.  With $a_{bs}^{2} = |a_{23}|^{2} + |a_{32}|^2$ 
and ignoring terms of $\ord{m_s/m_b}$, we get
\beq
\frac{{\rm Br}(B \to X_{s}\, \phi)}{{\rm Br}(B \to X_{c} \, e {\bar \nu_e})}  = 
\frac{6 \pi^2 a_{bs}^2 }{\Lphi^2 G_F^2}
\frac{(m_s/m_b^3) (1 - m_\phi^2/m_b^2)^2}{|V_{cb}|^2 f(m_c/m_b)}
\label{treeform}
\eeq
For $m_\phi \sim 1$~GeV and $\Lphi \sim100$~TeV, we obtain
\beq
{\rm Br}(B \to X_{s}\, \phi)  \sim 10^{-2} \; \; \;\;({\rm Tree}), 
\label{treenum}
\eeq
which is much larger than the loop effect in Eq.~(\ref{loopnum}).  Hence, 
we will focus on the much larger tree-level bulk effect for the rest of our discussion.   

For $1~{\rm GeV} \lsim m_\phi \lsim 4$~GeV, which is in the natural WTCM range, 
the $b$ quark decays are on-shell and perturbative QCD can be used to calculate 
hadronic decay rates of $\phi$.  
For $3.7~{\rm GeV}\lsim m_\phi \lsim 4$~GeV,
$\phi$ decays mostly into $\tau^+ \tau^-$ and $c {\bar c}$.  For
$m_\phi \lsim 3.7$~GeV, $\mu^+\mu^-, {\bar s} s$, and $g g$ are the most relevant 
decay modes.  We find for the partial widths
\beq
\Gamma_{\mu^+\mu^-} : \Gamma_{{\bar s} s}: \Gamma_{gg} \simeq 
x_\mu ^2 m_\mu^2 : 3 \,x_s^2 m_s^2 : \left(\frac{b_4 \alpha_s}{2\pi}\right)^2 \! m_\phi^2\, ,
\label{phimodes1}
\eeq
where $m_s \simeq 104$~MeV, $b_4 = 25/3$ is the QCD $\beta$-function
for 4 flavors, and $\alpha_s/\pi \simeq 0.1$.  Here $x = (c_L +
c_R)_f$, where $f=\mu, \, s$, and $c_{L,R}$ parametrize the
localization of $f$ left- and right-handed chiralities; 
we typically expect $x \approx 1$.  Thus, in this range of $m_\phi$, the radion 
nearly always decays into gluons.  We get the branching
fraction
\beq
{\rm Br}(b\to s\, \phi\to s \,g g)  \sim 10^{-2} \;\; \;\;({\rm WTCM}), 
\label{sggWTCM}
\eeq
which compared to that expected from the SM \cite{Simma:1990nr,Liu:1989pc}
\beq
{\rm Br}(b\to s \,g g)  \sim 10^{-3}\;\; \;\;({\rm SM})
\label{sggSM}
\eeq
implies
\beq
|a_{bs}|/\Lphi \lsim (10^4~{\rm TeV})^{-1}\, .
\label{bound1}
\eeq
The lifetime of $\phi$ is estimated to be 
\beq
\tau_\phi \sim \frac{32 \pi^3\Lphi^2}{b_4^2 \alpha_s^2 m_\phi^3} 
\sim \! 10^{-12}{\rm s} 
\left(\frac{\Lphi}{10^2~{\rm TeV}}\right)^2 \!
\left(\frac{1~{\rm GeV}}{m_\phi}\right)^3,
\label{tauphi}
\eeq
roughly corresponding to a displaced vertex of
$\ord{0.3}$~mm that could be a distinct signature.  
For $2 m_\pi \lsim m_\phi \lsim 1$~GeV we are in the  
non-perturbative regime and do not have reliable estimates.  
However, the main signal is expected to be $X_s \pi\pi$, with displaced $\pi \pi$
vertices.

Over the range $2 m_\mu< m_\phi < 2 m_\pi$, the 
partial widths will be given by
\beq
\Gamma_{\mu^+\mu^-} : \Gamma_{\gamma \gamma} \simeq 
m_\mu^2 :  m_\phi^2/(kL)^2.
\label{phimodes2}
\eeq
With $m_\phi \approx 2 m_\mu$ and $k L\simeq
30$ in the WTCM, the di-muon final state dominates in
this range.  Experimental data \cite{PDG} give 
\beq
{\rm Br}(B \to s \,\mu^+ \mu^-) = 
(4.3 \pm 1.2)\times 10^{-6} \;\; ({\rm Data}).
\label{Exp}
\eeq   
Demanding that the effect from $\phi$ be within 
the error on this measurement, we then get  
\beq
|a_{bs}|/\Lphi \lsim (2\times 10^5~{\rm TeV})^{-1}\,.
\label{bound2}
\eeq
For radion masses $15~{\rm MeV}\lsim m_\phi
\lsim 2 m_\mu$, the dominant decay channel is into di-photons and 
$\tau_\phi \gsim 10^{-6}$~s .  Hence, the signal will be $b \to s E\!\!\!\!\!/$, for this range of 
$m_\phi$;  we find that 
the experimental bounds on $B \to X_s \, \nu {\bar \nu}$ \cite{PDG} do not yield a 
severe constraint on the model parameters.  

Finally, we would like to mention that for $m_\phi \lsim
30$~MeV, $\phi$ is light compared to the supernova core temperature 
$T\sim 30$~MeV  and its emission can over cool the star.  Here, 
bounds from SN 1987A, based on energy loss in nucleon $N$ scattering $N N \to N N \phi$,  
imply $\Lphi \gsim 10^6$~TeV, corresponding to $m_\phi \lsim
100$~keV.  Hence, roughly, the interval 
$0.1~{\rm MeV}\lsim m_\phi \lsim 30$~MeV (corresponding to 
$10^6 {\rm~TeV} \gsim \Lphi \gsim 10^4$~TeV) is disfavored by 
the SN 1987 A data. 

In summary, the WTCM provides a mechanism to obtain the weak scale from a 
scale $\Lphi \sim100$~TeV.  This model typically predicts $m_\phi \lsim 4$~GeV, 
which suggests that $B$-decays may provide signals of the light radion.  We focused on 
the process $b\to s \phi$.  Here, the tree-level FCNC effects from non-universality dominate 
over loop-induced contributions.  We derived a number of bounds on WTCM parameters, 
using $B$-decay constraints and pointed out that the subsequent $\phi$ decays can lead to 
measurable displaced vertex signals.  We briefly discussed SN 1987A constraints 
and estimated that they disfavor the mass range $0.1~{\rm MeV}\lsim m_\phi \lsim 30$~MeV.

%%%%%%%%%%%%%%%%%%%%%%%%%%%%%%%%%%%%%%%%%%%%%%%%
%% BACKMATTER
%%%%%%%%%%%%%%%%%%%%%%%%%%%%%%%%%%%%%%%%%%%%%%%%

\begin{theacknowledgments}

We thank the SUSY 2009 organizers.  Work supported by the US Department of
Energy under Grant Contract DE-AC02-98CH10886.

\end{theacknowledgments}

\bibliographystyle{aipproc}   % if natbib is available

\bibliography{susy09}

\end{document}